\documentclass[preprint,aps]{revtex4}

\usepackage{graphicx}
\usepackage{booktabs}
\usepackage[usenames, dvipsnames]{color}
\usepackage{multirow}
\usepackage{setspace}
\usepackage{rotating}
\usepackage[
    pdftex, plainpages = false, pdfpagelabels, 
    pdfpagelayout = useoutlines,
    bookmarks,
    bookmarksopen = true,
    bookmarksnumbered = true,
    breaklinks = true,
    linktocpage=all,
    pagebackref=false,
    colorlinks = true,
    linkcolor = BrickRed,
    urlcolor  = blue,
    citecolor = BrickRed,
    anchorcolor = green,
    hyperindex = true,
    hyperfigures
]{hyperref} 

\newcommand{\comment}[1]{}

\newcommand{\sgn}{\mathrm{sign}}

\usepackage{url}
\usepackage{hyperref}

\begin{document}
\title{\href{http://www.necsi.edu/research/economics/bearraid.html}{Evidence of market manipulation in the financial crisis}\footnote{A report on preliminary results from this work was transmitted to the House Financial Services Committee and sent by Congressman Barney Frank and Congressman Ed Perlmutter to the SEC on May 25, 2010.}}
\author{
   \href{http://vedantmisra.com}{Vedant Misra}, \href{http://marcolagi.com}{Marco Lagi}, and \href{http://necsi.edu/faculty/bar-yam.html}{Yaneer Bar-Yam}} \thanks{\textnormal{ Corresponding author: \href{mailto:yaneer@necsi.edu}{yaneer@necsi.edu}}}
\affiliation{\textup{\href{http://www.necsi.edu}{New England Complex Systems Institute}\\
   238 Main Street Suite 319, Cambridge, Massachusetts 02142, US
}}
\date{\today}


\begin{abstract}
\begin{spacing}{1.39} 
We provide direct evidence of market manipulation at the 
beginning of the financial crisis in November 2007. The type of market
manipulation, a ``bear raid," would have been prevented by a regulation that was
repealed by the Securities and Exchange Commission in July 2007. The
regulation, the uptick rule, was designed to prevent market 
manipulation and promote stability and was in
force from 1938 as a key part of the government response to the 1929 market crash
and its aftermath. On November 1, 2007, Citigroup experienced an unusual
increase in trading volume and decrease in price. Our analysis of financial
industry data shows that this decline coincided with an anomalous increase in
borrowed shares, the selling of which would be a large fraction of the
total trading volume. 
The selling of borrowed shares cannot be explained by news events as there
is no corresponding increase in selling by share owners. A similar number of
shares were returned on a single day six days later. The magnitude and
coincidence of borrowing and returning of shares is evidence of a concerted
effort to drive down Citigroup's stock price and achieve a profit, i.e., a bear
raid. Interpretations and analyses of financial markets should consider the
possibility that the intentional actions of individual actors or coordinated
groups can impact market behavior. 
Markets are not sufficiently transparent to reveal or prevent
even major market manipulation events. 
Our results point to the need for
regulations that prevent intentional actions that cause markets to deviate from
equilibrium value and contribute to market crashes. Enforcement actions, even if they
take place, cannot reverse severe damage to the economic system. 
The current ``alternative'' uptick rule which is only in effect 
for stocks dropping by over 10\% in a single day is insufficient. 
Prevention may be achieved through a combination of improved transparency through 
availability of market data and the original uptick rule or other transaction process limitations.  
\end{spacing}

\end{abstract}
\maketitle


\section{Introduction to Bear Raids and Market Manipulation}\label{Introduction}

On July 6, 2007, the Securities and Exchange Commission (SEC) 
repealed the uptick rule, a regulation that was specifically designed to prevent market 
manipulations that can trigger market crashes. While it is widely accepted that the causes
of the crash that began later that year were weaknesses in the mortgage market and financial 
sector, the close proximity of the repeal to the market crash suggests that market manipulation 
may have played a role.
 
Here we present quantitative evidence of a major market manipulation, a ``bear
raid," that would not have been possible if the uptick rule were still in force. 
The timing of the bear raid, in autumn 2007, suggests that it may have
contributed to the financial crisis. Bear raids are an illegal market
strategy in which investors manipulate stock prices by collectively selling
borrowed shares. They profit by buying shares to cover their borrowed positions at
a lower price. While bear raids are often blamed for market events, including
financial crises ~\cite{nakedshorts, soros:bear}, this paper is the first to 
demonstrate the existence of a specific bear raid.

The sale of borrowed shares, called short selling, is a standard form of market
trading. Short sellers sell borrowed shares, then buy them back later and return
them to their owners. This practice yields profits when prices decline. In a bear
raid, investors engage in short selling with the addition of market manipulation. 
Instead of profiting from a natural decline in the fundamental value of a company
stock, the executors of a bear raid themselves cause the price to decline. 
Large traders combine to sell shares in high volume, ``driving" the price down~\cite{Brunnermeier, Ferri}. 

A bear raid is profitable if other investors are induced to sell their shares
at the lower price. This may happen for two reasons: margin calls and panic.
Margin calls occur when brokerages force investors to liquidate their
positions. Investors who are confident in the rising price of a stock may buy
shares on borrowed funds, called ``buying on margin," using the value of the
shares themselves as collateral. When prices decline, so does the value of 
the collateral and at some point brokerages issue ``margin calls," requiring 
shares to be sold even though the owners would prefer not to. 
Panics occur when investors, fearing further losses, sell 
their shares. The executors of a bear raid profit from the price decline by 
buying back the shares they borrowed---``covering" their short positions---at 
the lower market price.

In the aftermath of the 1929 market crash, Congress created the Security and
Exchange Commission (SEC). Recognizing the dangers of short selling, Congress
specifically required the SEC to regulate short
selling~\cite{SEC1934}. The regulation that was instituted in 1938, the
uptick rule, states that borrowed shares may only be sold on an ``uptick"---at
a price that is higher than the immediately preceding price. The rule was
designed to limit the intentional or unintentional impact of short selling in
driving prices down, and specifically to prevent bear raids. The uptick rule
was repealed in July, 2007 by the SEC on the basis of arguments that 
markets were transparent and no longer needed the protection of the 
uptick rule~\cite{repeal1}. SEC claims that the uptick rule had no significant
effect on market stability, even in absence of specific manipulation, have been
refuted ~\cite{NECSI3, NECSI4, NECSI5}. Our results implying a bear raid in 
November 2007 contradict the assertion of market transparency.

Our evidence points to a bear raid on the large financial services company
Citigroup. On November 1, 2007, Citigroup's stock experienced an unusual
increase in trading volume and decrease in price. To analyze this event, we
studied financial industry short trading data (see Appendix A), which reveal 
the total number of borrowed shares (short interest) at the end of each trading day. Using these
data, we show that the increase in trading volume on November 1 coincides with
an increase in borrowed shares. Six days later, a comparable number of short
positions were closed during a single trading day. News events to which these
events might normally be attributed cannot account for the difference between
trading in borrowed shares and trading by owners of shares. The magnitude and
coincidence of short activity is evidence of a concerted effort to drive down
Citigroup's stock price and achieve a profit, i.e., a bear raid.

\section{Citigroup on November 1 and 7, 2007}

On November 1, 2007, Citigroup experienced large spikes in short selling and
trading volume. The number of borrowed shares---short interest---increased by 
approximately 130 million shares to
3.8 times the 3-month moving average. The total trading volume jumped from 73
million shares on the previous day to 171 million shares, 3.7 times the 3-month
moving average. The ratio of the increase in short positions to volume was
0.77. This is the fraction of the total trading that day that may be attributed
to short positions held until market closing. The total value of shares
borrowed on November 1 was approximately \$6.07 billion. Adjusted for the
dividend issued on November 1, 2007, Citigroup stock closed on November 1 down
\$2.85 from the previous day, a drop of 6.9\%.

The number of positions closed on November 7, 202 million, was 53\% larger than
the number opened on November 1. The short interest before the increase on
November 1 and after November 7 are virtually identical, the larger decrease
corresponding to an additional increase in short interest between these dates.
The mirror image one-day anomalies in short interest change suggest that the
two are linked. We can conservatively estimate the total gain from short
selling by multiplying the number of short positions opened on November 1 by
the difference between the closing price on November 1 and closing price on
November 7 (\$4.82), which yields an estimated gain for the short sellers of \$640 million. 

The total decrease in short interest on November 7 exceeds the total trading
volume on that day, 121 million, by 82 million shares. This indicates that the
reported decrease in borrowed shares is not fully accounted for by recorded
trading on the markets. The difference may result from off-market transfers,
which may be advantageous to short sellers in not causing the price to
increase. Alternatively, despite the usual coincidence of borrowing and
selling, this may be due to shares that were borrowed and returned without
being sold short. Further investigation of transaction data is necessary to
explain the difference in returned shares and trading volume. 

\begin{figure}[tbh]
    \centering
    \href{http://www.necsi.edu/research/economics/bear-raid_vol-price-si_large.pdf}{\includegraphics[width=.9\textwidth]{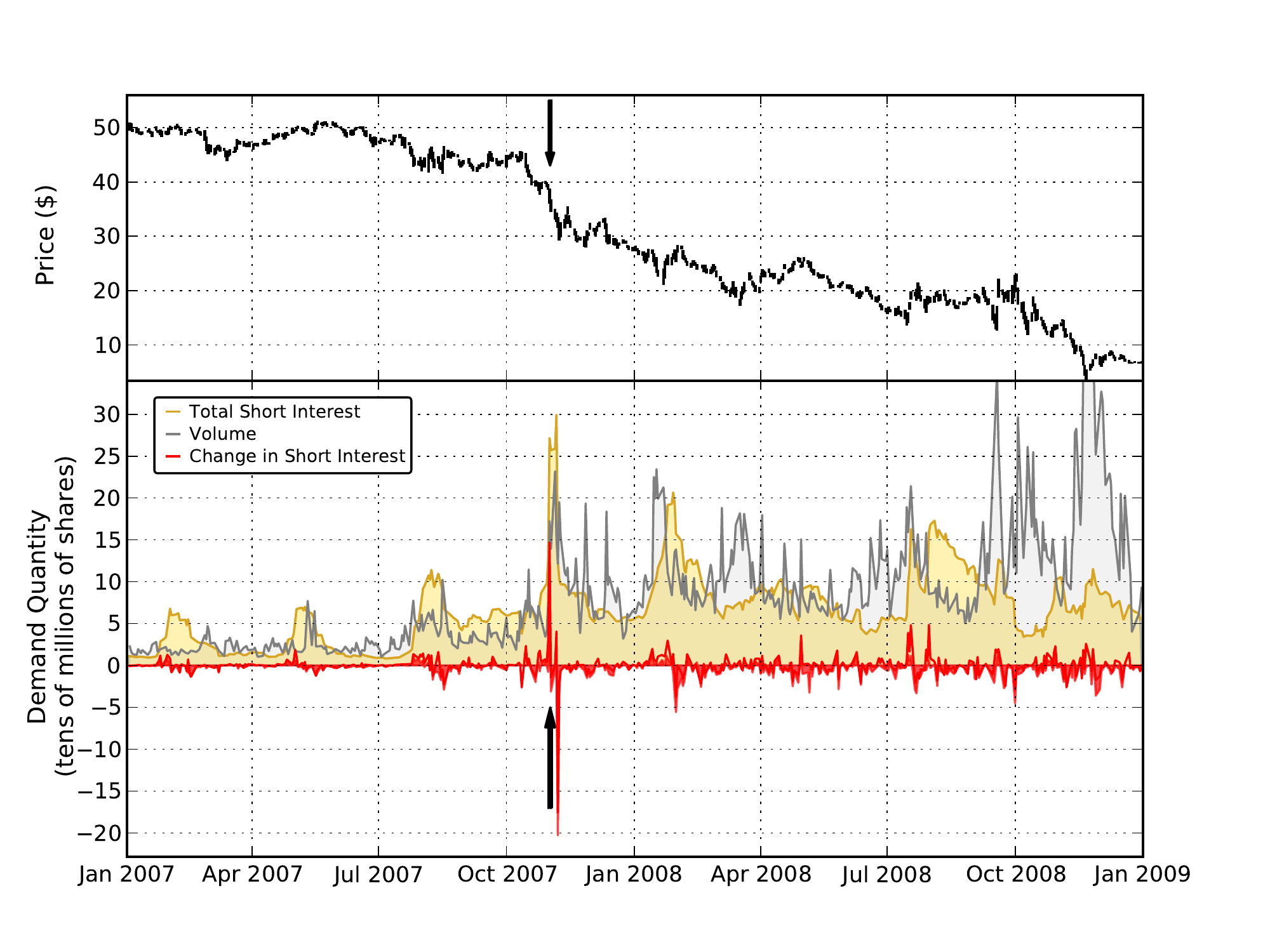}}
    \caption{Market activity for Citigroup over a two-year period starting
    January 1, 2007. Top panel shows vertical bars for the daily high and low stock 
    price. Lower panel shows total short interest (yellow), trading volume (gray),
    and daily change in short interest (red).
    Arrows indicate November 1, 2007~\cite{data}.
    \label{fig:citi_big}}
\end{figure}

Figure~\ref{fig:citi_big} shows daily stock price, volume, and short sale data
for Citigroup over a two-year period starting January 1, 2007. Short sale data
includes short interest---the number of shares borrowed at the end of each
\begin{figure}[tbh]
    \centering
    \href{http://www.necsi.edu/research/economics/bear-raid_vol-price-si_small.pdf}{\includegraphics[width=.9\textwidth]{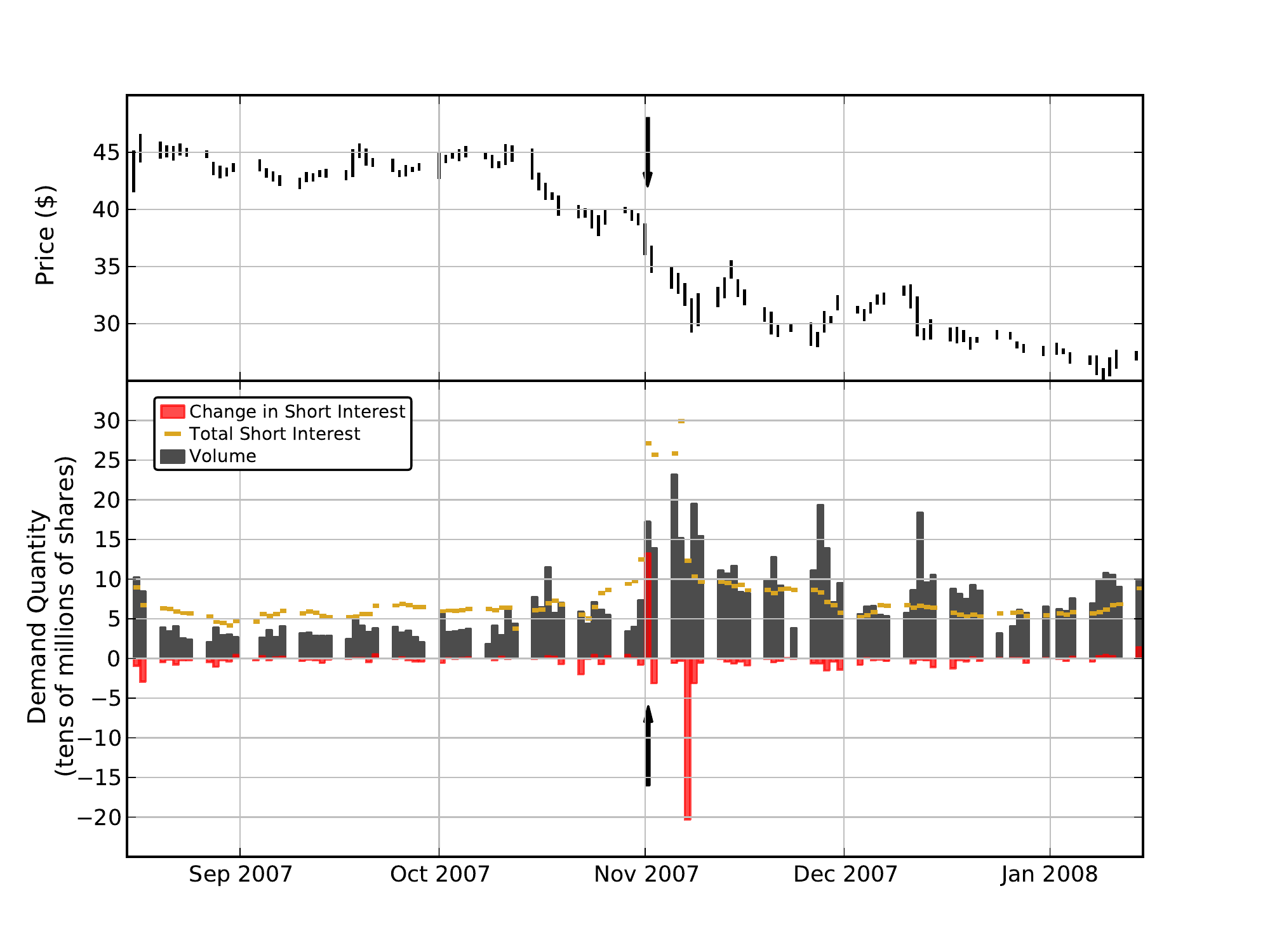}}
    \caption{Market activity for Citigroup over a five-month period starting on 
    August 15, 2007. Top panel shows bars for daily high and low stock price 
    (adjusted for dividends). Lower panel shows daily change in short interest 
    (red bars), total short interest 
    (yellow lines), and trading volume (gray bars). Arrows indicate November 1, 2007~\cite{data}.
    \label{fig:citi_small}}
\end{figure}
day---and the daily change in short interest. During much of 2007-2009, the
daily change in short interest did not exceed a small fraction of the total
trading volume. The largest single-day increase in short interest occurred on
November 1 and is marked with arrows in Figure~\ref{fig:citi_big}.
Figure~\ref{fig:citi_small} shows an enlarged view of the period around that
date.

\begin{figure}[tb]
    \centering
    \href{http://www.necsi.edu/research/economics/bear-raid_scatter.pdf}{\includegraphics[width=.8\textwidth]{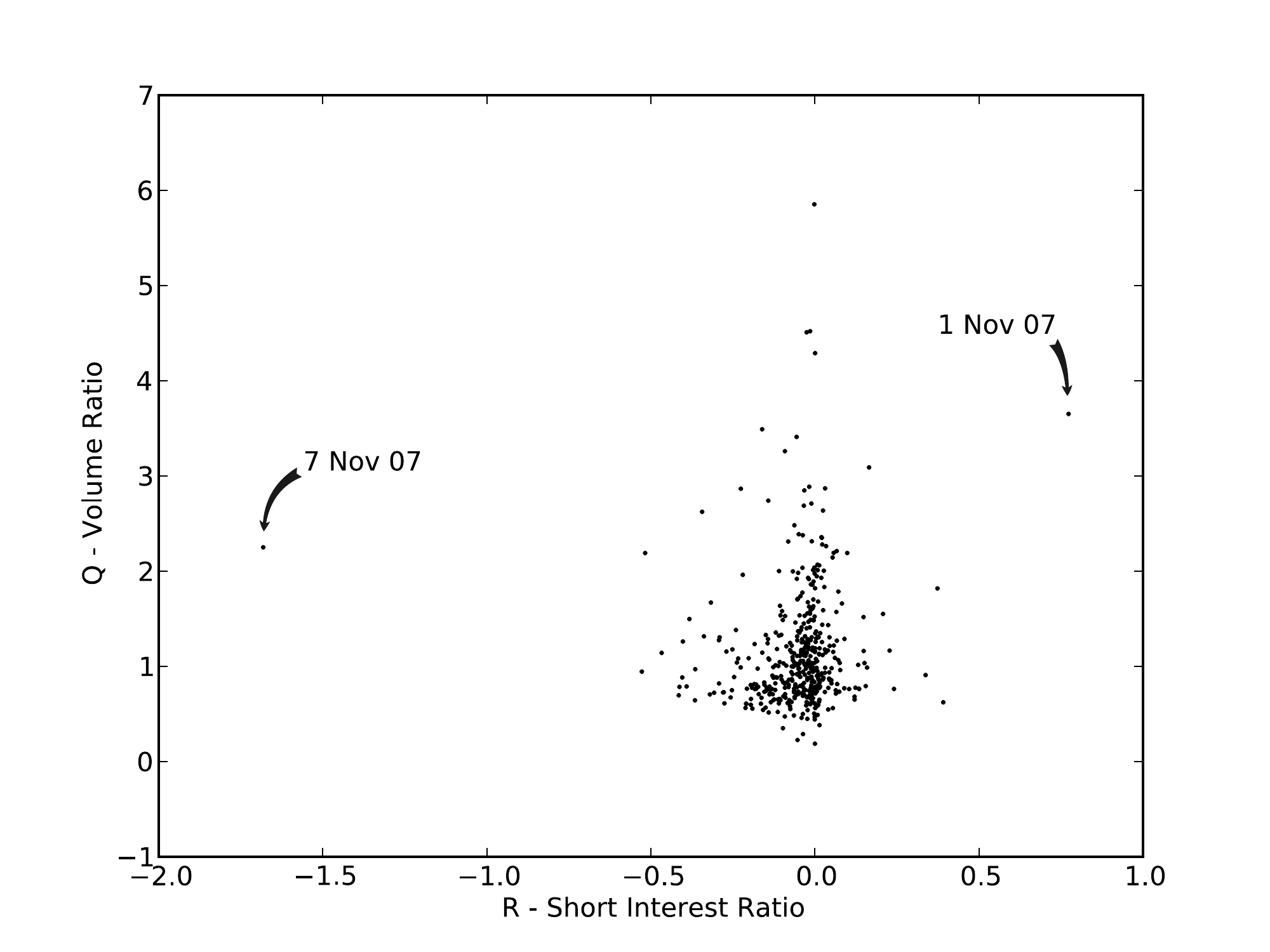}}
    \caption{Scatter plot of the daily volume of trading divided by the three month prior average (volume ratio), and the increase in number of borrowed shares divided by the volume (short interest change ratio), for Citigroup over a two-year period starting January 1, 2007. Arrows indicate Citigroup on 1 November 2007 and 7 November 2007. These two points are well outside of the behavior of daily events even during the period of the financial crisis in late 2007 and throughout 2008. The two measures are described in Appendix A.
    \label{fig:scatter}}
\end{figure}

In Appendix B we analyze quantitatively the probability of the events on
November 1 and November 7. Often
probabilities are estimated using normal (Gaussian) distributions that 
underestimate the probability of extreme events
(``black swans") that are better represented by long-tailed distributions~\cite{Mantegna, taleb}. 
We directly fitted the long tails of the distributions and estimated
the probability of the events based upon these tails to be $p =
2\cdot10^{-5}$ and $8\cdot10^{-9},$ respectively. Given 250 trading days in
a typical year, it
would take on average 200 years and 500 thousand years, respectively, to
witness such events. Moreover, the probability of these two events occurring 6
days apart is $p = 1\cdot10^{-12}$, corresponding to 4 billion years, comparable to the age of the Earth. 
Figure~\ref{fig:scatter} shows that these events are outside the general behavior of the market.
We emphasize that our estimates of the probabilities of these events reflects the higher 
probabilities of extreme events in long-tailed distributions.

Changes in investor behavior are often explained in terms of specific news
items, without which it is expected that prices have no reason to change
significantly~\cite{Fama, FamaFrench}. The press attributed the drop of
Citigroup's stock price on November 1 to an analyst's report that
morning~\cite{Dealbook, Rosenbush}. This report, by an analyst of the Canadian
Imperial Bank of Commerce (CIBC), downgraded Citigroup to ``sector
underperform"~\cite{Whitney}. Any such news-based explanations of investor
behavior on November 1 (similarly for November 7) would not account for the
difference in behavior between short sellers and other investors. Under the
assumptions of standard~\cite{FamaFrench} capital asset pricing models, all
investors act to maximize expected future wealth~\cite{Sharpe}, and should
therefore respond similarly to news. Furthermore, it has been shown empirically
that the ratio of short sales to total volume remains nearly constant, even
around news events~\cite{Engelberg}. In the literature, analysis of the
residual small differences in the behavior of short and long investors has been
interpreted to indicate that short sellers have an informational
advantage or that short sellers are able to
anticipate lower future returns~\cite{Engelberg, asquith, senchack, boehmer, Desai},
rather than cause them. Still, these studies do not show that large differences
in trading generally occur between short and long sellers. Thus, the existence
of such a difference is indicative of specific trader action.

Our evidence points to a bear raid during a period of financial stress~\cite{Longstaff, Caballero} to which the
Federal Reserve Bank responded in August 2007 by announcing that they would be
``providing liquidity to facilitate the orderly function of markets" because
``institutions may experience unusual funding needs because of dislocations in
money and credit markets"~\cite{fedcredit}. 
Shortly afterwards, the Dow Jones Industrial Average
achieved its historical peak---14,167 points on October 9---three weeks
prior to November 1, the date our evidence suggests a bear raid occurred.
Bear raids may have long-term price impact if decision makers infer investor
confidence from price movements and act on that basis~\cite{Goldstein, Khanna}.
Citigroup CEO Charles Prince's resignation on November 4 after an emergency
board meeting~\cite{Wilchins} may
reflect such an effect. The months after November 1 saw the beginning of the
stock market turmoil of 2008-2009 as well as many significant events of the financial
crisis, such as the purchase of Bear Stearns by JP Morgan Chase in March 2008
and the bankruptcy of Lehman Brothers in September 2008. 


\section{Conclusions and policy implications}\label{Conclusions}

The 2007--2011 financial crisis resulted in widespread economic damage and
introduced questions about both our understanding of economic markets and about
the practical need for regulations that ensure market stability. The Financial
Crisis Inquiry Commission (FCIC) reported that over 26 million Americans were
unemployed or underemployed in early 2011, and that nearly \$11 trillion in
household wealth evaporated. Moreover, the FCIC concluded that the crisis was
avoidable and was caused in part by ``widespread failures in financial
regulation and supervision [that] proved devastating to the stability of the
nation's financial markets" \cite{FCIC}. Regulatory changes that preceded the
financial crisis include the June 2007 repeal of the uptick rule, which was
implemented in 1938 to increase market stability and inhibit
manipulation~\cite{pessin,NECSI3, NECSI4, SEC1934, repeal1}. 

Within the resulting deregulated environment, it is still widely believed that the crisis was caused 
by mortgage-related financial instruments and credit conditions, and that individual traders
did not play a role~\cite{MahHui,Blanchard,Acharya,Hellwig}.
Our analysis demonstrates that manipulation may 
have played a key role. Methods for detecting manipulation and its effects are
necessary to both inform and enforce policy.

When the SEC repealed the uptick rule on July 6, 2007, one of its main claims was that 
the market was transparent, and that such regulations were not needed to 
prevent market manipulation~\cite{repeal1}. Our results suggest
that, not long after the uptick rule was repealed, 
a bear raid may have occurred and remained undetected and unprosecuted.
Our analysis reinforces claims that lax regulation was an integral
part of the financial crisis~\cite{FCIC}.

In response to requests for reinstatement of the uptick rule after the financial crash, 
the SEC underwent extended deliberations and finally implemented an alternative uptick rule, 
which allows a stock to fall by 10\% in a single day before limitations on short selling apply~\cite{SECcb}.
This weaker rule would not have affected trading of Citigroup on November 1, 2007, as its minimum price
was just 9\% lower than the close on October 31. Subsequent day declines until November 7 were also 
smaller than 10\%.

The existence of a major market manipulation should motivate changes in market
models, analysis, regulation and enforcement. In particular we conclude that: 
\begin{itemize}
\item Large traders may have a significant influence on the market. Scientific analysis
and models should recognize the role of large traders and consider both past
events and potential future events they may cause. For example, market time
series analysis that does not specifically consider the effect of manipulation
may be unable to discover it, because manipulation events may not manifest in
averages and distributions that are usually considered.
\item Improved access to data can enable the detection of market manipulation. This
would foster transparency in the markets, which has been lauded but not realized.
Regulatory agencies should mandate the 
increased availability of relevant data for the detection of manipulation. If these
data cannot be made available in real-time or for public use, they may be
provided with time delays or only for scientific use. Data of importance
include not only the opening of short positions but also their closing, as
aggregate short sale activity cannot be determined when only opening trade data
are available. These data should be made available at the transaction level.
\item Current legislation, which focuses on retroactive penalties, is
ineffective due to the discrepancy between the timescale of enforcement
response and that of market manipulation. Severe failures in the financial
system may include cascading global market crises and numerous takeovers and
bankruptcies, making the disentanglement of individual events difficult if not
impossible. Regulatory agencies should adopt preventive measures such as 
the uptick rule, which would be more effective
than punitive ones. The uptick rule was designed to minimally restrict trader's
actions while simultaneously providing underlying stability for the financial
system and inhibiting particular forms of manipulation, including bear raids. 
\item The limitations of our data prevent definitive conclusions about
individual events or their attribution to individual investors. Enforcement agencies 
should perform investigations into specific candidate events, including the 
candidate event we identified on November 1, 2007.
\item Until effective regulations and enforcement are in place, market price
changes may not reflect economic news. They may reflect market manipulation. 
\end{itemize}

The complexity of financial markets and their rapid dynamics
suggest that data analysis and market models are increasingly necessary for
guiding decisions about setting market regulations and their enforcement~\cite{Haldane,
Johnson, Lux2}. Independent of the role it may play in financial crises,
understanding market manipulation may be important for characterizing market
dynamics. Recent decades have seen significant advances in financial market
theory, including the mean-variance portfolio theory~\cite{Markowitz}, the
capital asset pricing model~\cite{Sharpe}, arbitrage pricing
theory~\cite{Ross}, and the theory of interest rates~\cite{Cox}. However, the
financial crisis and anomalous events such as ``flash crashes" \cite{Khandani}
demonstrate limitations in existing approaches. More recent efforts seek to
explain market phenomena via methods such as agent-based modeling~\cite{
Gangoffourmodel, Lux, Samanidou, LevyLevy, ContBouchaud, DonangeloSneppen}
and analysis of the long-tailed distributions of price fluctuations~\cite{Gabaix,
Liu, Mantegna, Sornette, SolomonLevy}. While these methods have
been successful in describing some aspects of market behavior, they generally
do not consider the impact of individual traders who have the ability to
significantly impact the market~\cite{AllenGale, AllenGorton, Jarrow, Kyle,
BenabouLaroque, KumarSeppi, Aggarwal}. Current approaches,
whether analytical or statistical, may not reveal isolated---or even
frequent---instances of trader influence. 

Among the possible forms of individual trader influence, intentional
actions---including manipulation---are of particular relevance, as they
undermine the role of markets in setting prices so as to reflect economic
value. Market manipulation is illegal under Section 10 of the Securities
Exchange Commission Act of 1934~\cite{SEC1934}. Some forms of manipulation are
well documented, including indirect price manipulation through the generation
of false news~\cite{pumpanddump}. Direct price manipulation through market
transactions is also commonly thought to occur~\cite{AllenGale,nakedshorts,
soros:bear}, but methods for its detection that are based on statistical
analysis~\cite{Minenna, Cholewinski} are limited by their inability to
independently account for news events and other anomalies. No direct evidence
of recent price manipulation has been presented based upon these methods.

The timing of the event we identified raises questions about the potential role
it may have played in the financial crisis. Understanding the wider impact of
such an event requires that we consider the vulnerability of the overall
market. 

Whereas a highly stable system is not vulnerable to any but the largest
impacts, a vulnerable system can be destabilized by much smaller 
shocks~\cite{DeBandt, Buldyrev}. This is a general aspect of the behavior of
complex interdependent systems, not just of financial markets. Specific events
can have large effects if the underlying physical, biological or social system
is vulnerable. For example, while mass extinctions have been shown to coincide
with meteor strikes~\cite{Schulte}, underlying vulnerabilities are thought to 
contribute to the severity of extinction events~\cite{Arens}.
Similarly, market manipulation during a period of instability and high
interconnectedness, such as before the financial crisis~\cite{Longstaff,
Caballero, NECSI1}, may exacerbate or even trigger a collapse. 
The financial system can be expected to exhibit this general property of complex systems,
in which the coincidence of underlying vulnerability and extreme events
can trigger crises.

We thank Yves Smith and Matt Levine for helpful comments. This work was supported by the New England Complex Systems Institute. 


\section*{Appendix A: Methodology: Data and Event Detection}\label{Methodology}

It is generally difficult to characterize the investments of individual
traders, especially for short positions. Unlike those who own large stakes in
companies, those with large short positions are not required to report their
holdings~\cite{sched13d}. Short
interest data is publicly available by ticker symbol at two-week intervals for
a rolling 12-month period~\cite{nasdaqSI}. This time resolution
is too low to detect the bear raid candidate we will describe, and does not
include historical data for the period of the financial crisis. The recent
availability of off-market transaction systems that enable large volume
transactions, such as crossing networks \cite{mittal, harris}, makes it
difficult, if not impossible, to trace intentional large short sale
transactions using market data. A short sale transaction between cohorts on a
crossing network may allow one trader to execute a short sale while the other
trader accumulates a long position. This long position can then be sold on the
open market without leaving a signature of its short sale origins.

Our study is based on industry data on daily securities lending. While this
data does not identify the individuals borrowing the shares, the time
resolution proved sufficient to provide evidence of a bear raid.

We obtained price and volume data from Thomson
Reuters Datastream. Short interest data was obtained from Data
Explorers and included a daily record of the value and quantity of loaned
securities as reported by brokerages. These included separate time series for
the total number of borrowed securities (total demand quantity) and for daily
incremental changes in the number of  borrowed shares.  Daily incremental
changes were approximately given by day-to-day differences in  total demand
quantity, with small corrections arising from the addition and  removal of
reporting organizations from the data set. The reconstruction of short selling
data from security lending data is an inexact process, because borrowed
securities may be used for purposes other than short selling, including tax
arbitrage, dividend arbitrage, and merger arbitrage.   Furthermore, reported
data may be incomplete, because not all lenders supply data to industry data
providers. Nevertheless, because short selling is the predominant reason for securities
lending, securities lending is a reasonable proxy for short selling~\cite{Faulkner,
DataExplorersReport}. 
We also were able to eliminate the possibility of the most likely alternative 
explanation to a bear raid, dividend arbitrage, as described in Appendix C. 

The signature of a successful bear raid is an anomalous spike in the number of
shares of a company's stock that are sold short, followed by a price decline,
then a corresponding large spike in the number of positions that are
covered---a decrease in the number of short positions. A sufficiently large
increase in short selling would also increase the total volume of trades, so we monitored
also the total daily trading volume. 

We searched data for several prominent companies to identify candidate events,
and calculated two ratios, $R$ and $Q$, for each
trading day. $R$ is the ratio of the change in short interest to daily volume,
\begin{equation}
    R(t) = \frac{\Delta S(t)}{V(t)},
    \label{eqn:R}
\end{equation}
where $\Delta S(t)=S(t)-S(t-1)$ is the change in short interest, $V$ is trading volume, and
$t$ is the date. A large absolute value of $R$ indicates that a high proportion
of trading is accounted for by securities lending activity---that the volume
of borrowed shares was a substantial fraction of the total volume, and that
short sales might have affected the stock price. A high positive value
indicates that shares were borrowed, and a high negative value indicates short
covering. Note that if a large number of short positions were opened and closed
on the same day (i.e. an intraday bear raid), it would not be revealed by daily
short interest data. We cannot exclude the possibility of intraday bear raids
occurring during this period. 

$Q$ is the ratio of the trading volume to the three month moving average,
\begin{equation}
    Q(t) = \frac{V(t)}{\overline{V}(t)},
    \label{eqn:Q}
\end{equation}
where $\overline{V}$ is the prior 3-month (63 trading day) moving average of volume.
A value of $Q$ substantially greater than one indicates an anomalously high 
trading volume. The event we analyzed was identified by a high absolute value of $R$ 
and high value of $Q$, indicating that
the increase in borrowed shares was large in comparison to trading activity,
and that total trading activity increased dramatically. 

\section*{Appendix B: $R$ and $Q$ distributions}

In this appendix we present our analysis of the distributions of $R$ (the ratio
of the change in short interest to daily volume, see Eq. \ref{eqn:R}) and $Q$ (the
ratio of the trading volume to the three month moving average, see Eq. \ref{eqn:Q})
for Citigroup, from January 2007 through December 2008. The analysis allows us to obtain a
probabilistic estimate of the inherent likelihood of $R$ and $Q$ values for
each day, and in particular for the events on November 1 and 7, 2007.

\begin{figure}[tbp]
\href{http://www.necsi.edu/research/economics/bear-raid_R-right.pdf}{\includegraphics[width=.9\textwidth]{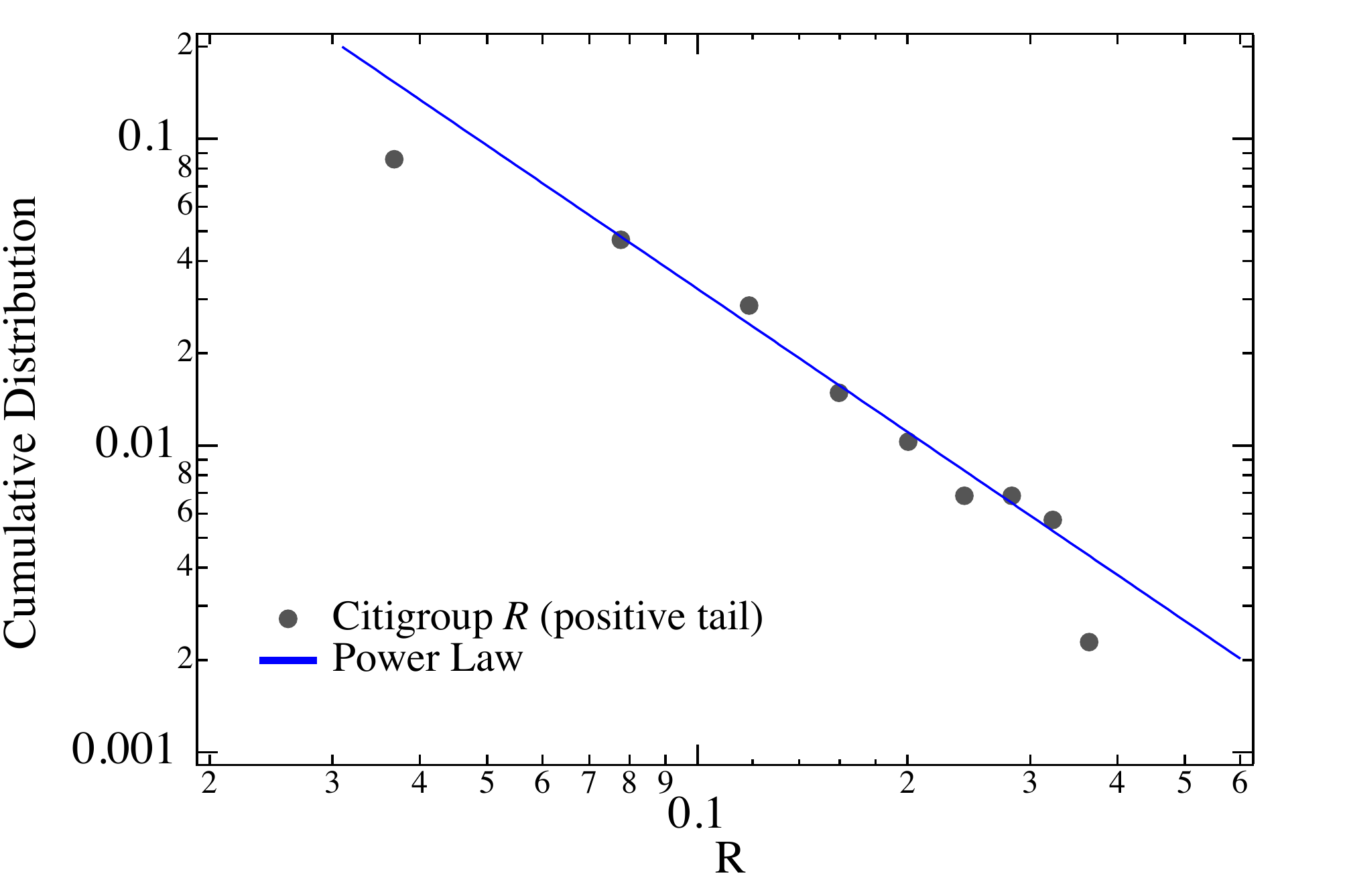}}
\href{http://www.necsi.edu/research/economics/bear-raid_R-left.pdf}{\includegraphics[width=.9\textwidth]{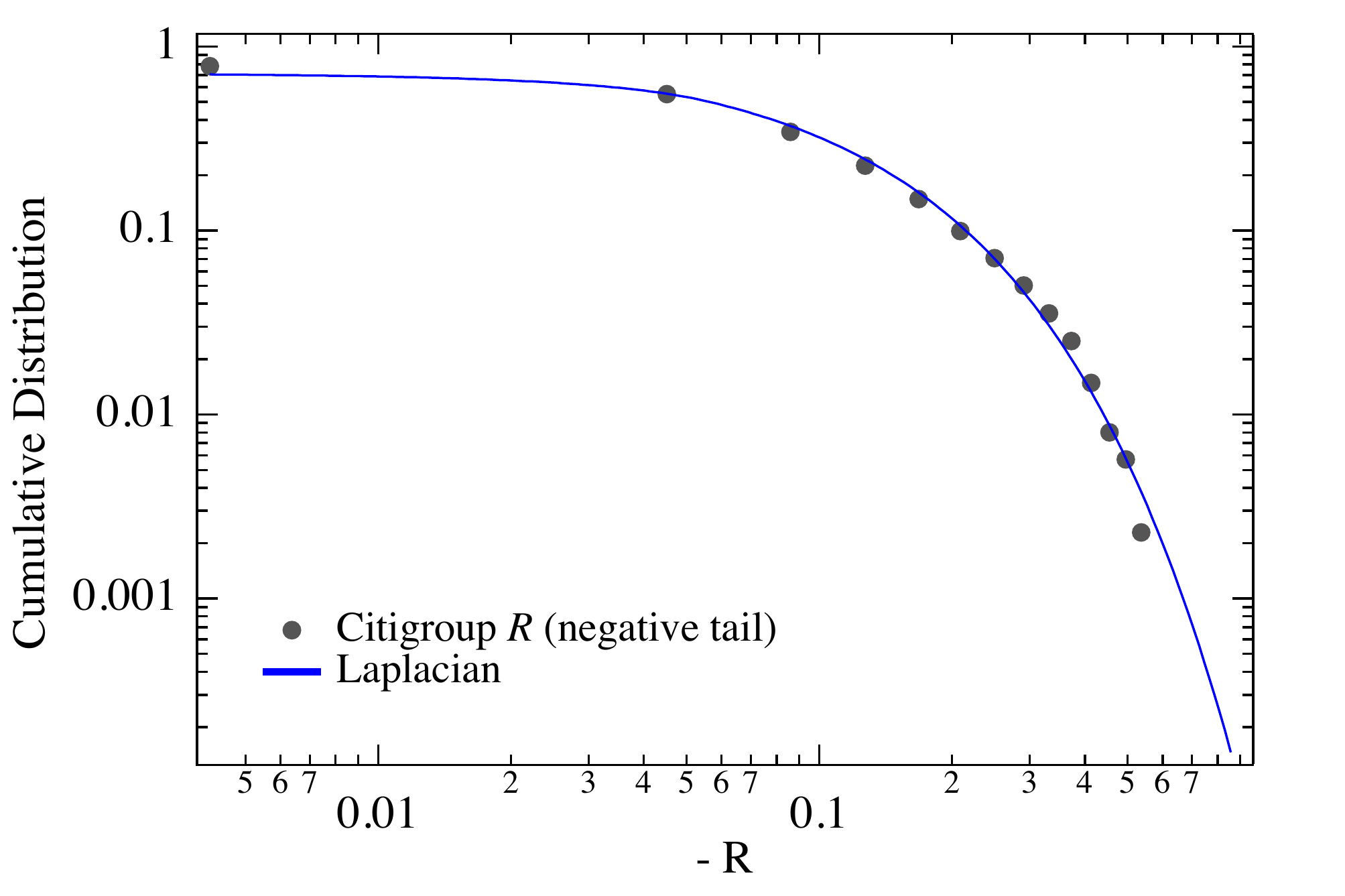}}
\caption{\textbf{Citigroup $R$ distribution} - Cumulative distribution
functions (CDF) of the short interest change ratio for Citigroup, for 2007 and 2008. 
\emph{Top panel}: Positive tail of the
distribution, blue line is the best fit power law (CDF$(R)$
$\sim R^\alpha$, with $\alpha=-1.35$). \emph{Bottom panel}: Negative tail of the 
distribution, blue line is the best fit Laplacian distribution (CDF$(R)$ $\sim
1+\sgn(R-\beta)(1-\exp(-|R-\beta|/\gamma))$, with $\beta=0.11$ and
$\gamma=0.048$).
\label{fig:R}}
\end{figure}

\begin{figure}[tbp]
\href{http://www.necsi.edu/research/economics/bear-raid_Q.pdf}{\includegraphics[width=.9\textwidth]{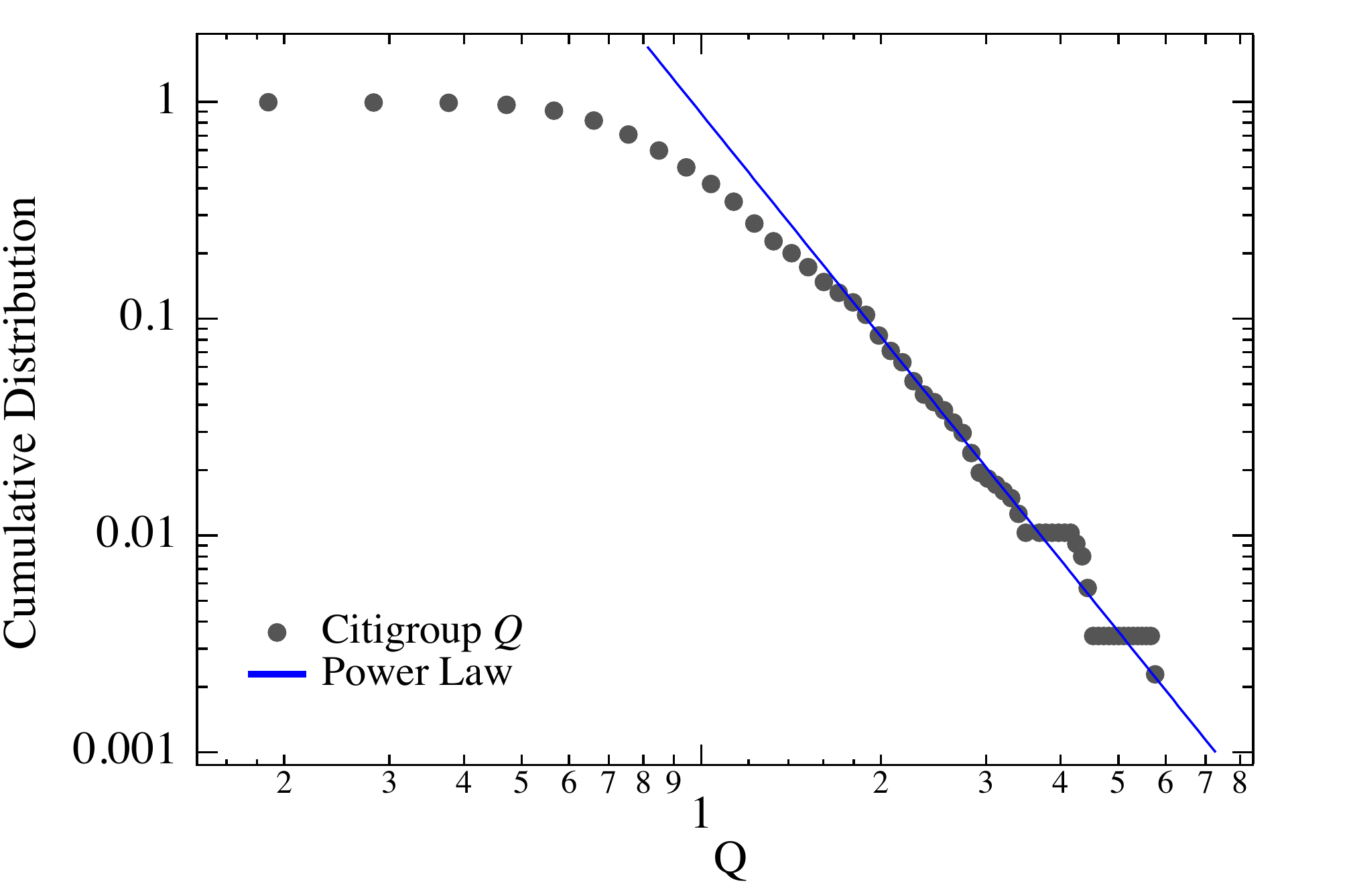}}
\caption{\textbf{Citigroup $Q$ distribution} - Cumulative distribution function (CDF) of the volume ratio for Citigroup for 2007 and 2008. Blue line is the best fit power law (CDF$(Q)$ $\sim Q^\alpha$, with $\alpha=-3.34$).
\label{fig:Q}}
\end{figure}

The positive and negative tail cumulative distributions for Citigroup for $R$ are plotted in
Fig. \ref{fig:R}. 
The two sides of the distribution behave differently: while
the positive tail follows a power law distribution (top panel), the negative
tail is well described by a Laplacian distribution (bottom panel). The distribution 
for $Q$, shown in Fig. \ref{fig:Q}, has a power law tail. November 1 and 7, 2007 
are omitted in the plots, but this does not affect the fitted distributions.
From the fitted distributions we extracted the expected probabilities of the two
events. 

\section*{Appendix C: Tests and Technical Notes}

We have tested a number of alternative explanations of the data:

\begin{itemize}

\item Is it possible that the borrowed shares were used to receive a
dividend payment, i.e. dividend arbitrage? 

Sometimes borrowing shares provides
benefits of dividends to the borrower rather than to the owner. In such cases
the borrower may not necessarily sell the shares short, which precludes a bear
raid.

The date on which shares were borrowed, November 1, was an ``ex-dividend"
date, i.e. a date on which ownership determines dividend payments. In order for borrowers to
receive the benefit of dividends they are required to hold the
shares at the prior day's closing. Thus, there was no dividend paid to shares
borrowed on November 1.

\item Is it possible that the reported dates for borrowed shares is delayed so that
the actual date of borrowing is a different date than what is reported (for
example, could it be reported on the date of settlement three days after a
market transaction)? 

We verified the agreement of reported borrowing and short selling date by
looking at the period of the short sale ban starting in September 2008. The dates of
the start and stop of borrowing coincide with the dates that they should for the
ban, which shows that there is no delay in reporting.

\item Does commercial market transaction data corroborate the short selling?

We have studied commercially available NYSE short selling data
\cite{nyseshortsales} from these dates, and found it to be unreliable because
the transactions reported are inconsistent with reported trade and quote data
\cite{nysetaqtrade} at the transaction level. Despite dialog with the NYSE
staff we have not received an explanation of the inconsistency. For the present
analysis, the inconsistency inhibits our efforts to use this data to
cross-validate the results in this report. More generally, it raises questions
about the reliability of market provided short sale data.

\item Is it possible that the analyst report downgrading Citigroup that 
morning was released in collusion with the bear raid?

We have no specific evidence, but such collusion would be consistent 
with strategies used by those who manipulate stocks ~\cite{pumpanddump, AllenGale, nakedshorts, soros:bear}.

\item Is it possible that those who engaged in the bear raid also used trading
in options to increase their profits by buying put or selling call options?

Our estimate of the profits made on the bear raid are conservative.

\item Is it possible that the large block trades on November 1 and 7
represented trading based upon information
that was not yet available to the public on November 1?

Our evidence suggests that a single individual or group of individuals traded a
large volume of borrowed shares on November 1 and November 7. If this
represented potentially illegal insider trading, the traders would have avoided
attracting attention. Neither the large trading volume nor the
abrupt price drop on November 1 at the opening of the market appear to be consistent
with a low-profile trading approach. The rapid price drop is also inconsistent
with the expected behavior of insider traders, which is to maximize profits by selling gradually to avoid
affecting prices until the negative news becomes public. Both the
large volume of trading and the rapid drop are consistent with trading intended
to affect prices, i.e. a bear raid. While the intentions of traders can only be
determined from a more detailed inquiry once those traders are identified, the
available information strongly supports a bear raid over the possibility of
insider trading per se. It is possible that traders with insider information 
chose to help matters along by performing a bear raid at the same time 
as they were trading on insider information.

\end{itemize}

\section*{Addendum: Additional Tests and Technical Notes}

Following the release of this paper, we were contacted by the NYSE with additional
information about the NYSE short selling transaction data \cite{nyseshortsales} described in
Appendix C. The new information enabled us to reconcile the short sale and trade 
data \cite{nysetaqtrade} by aggregating and shifting the times of multiple transfers to 
correspond with market transactions. There are residual issues with a small minority
of transactions that are being resolved, but these issues appear to be irrelevant to 
conclusions about the volume of trading. 

The additional information enables us to identify with some confidence the 
reported short sale volume on the NYSE on November 1 and other dates. 
The short sale volume is not unusual as a proportion of total volume, constituting  
about one quarter of the total volume on this market. NYSE transactions constituted 30\% of the total
market volume on November 1, 2007. This limits the volume of reported short selling on the markets, 
and diminishes the likelihood that the reported increase in borrowed shares was directly reflected in 
reported short sales.

Absent an alternative interpretation, if shares were sold in a way that concealed their 
origin as borrowed shares the data sets would be consistent. 
One method to achieve this, using ``short to buy" transactions,
was reported in Senate investigations of the Pequot Capital hedge fund in 2009  \cite{pequot}.
In this approach a single trader moves shares from one account to another, creating a 
short position in one and a long position in the other. Since there is no change in 
beneficial ownership, such transactions may be reported in a way that is not consistent 
with standard reporting requirements, resulting in share borrowing without a market record. 
Long positions created this way may be sold on 
any market without being identified as short sales, even though in doing so a net short 
position is created.

This method appears to have been developed to hide short selling at a time when the uptick rule was in effect.
Short to buy transactions require a close relationship with a broker dealer. The necessary 
access to market trading systems, called ``sponsored'' or ``direct market'' access, needed to perform 
the short to buy transaction is not available to most traders but constitutes a significant 
fraction of reported trading \cite{tradersmag,econwar}. 
Only recently, beginning in 2011, were brokers required to apply standard regulations to transactions 
of traders using sponsored access \cite{final_rule,sponsoredaccess}. Previously, non-compulsory 
self-regulation was in effect \cite{proposed_rule}. In the absence of oversight, market 
data may not properly record the volume of short selling. 

An explanation in these terms for the events in November of 2007 is also consistent with the 
observation that there was a larger volume of returned shares on November 7 than the
trading volume. In the ``short to buy'' scenario, residual positions can be closed through ``back
office'' transactions and may never be recorded on the market.

The new information we received implies that the sale of borrowed shares reflected in 
the increase in borrowed shares on November 1 and the corresponding decrease
on November 7 may have been done in a way that would not have been
prevented by the uptick rule. A more detailed inquiry into the means by which
such selling could have been done is beyond the current work. 

We thank Steven Poser and Wayne Jett for helpful discussions.

\end{document}